\documentclass[%
reprint,
superscriptaddress,
nofootinbib,
nobibnotes,
 amsmath,amssymb,
 aps,
pre,
]{revtex4-1}

\usepackage{graphicx}
\usepackage{dcolumn}
\usepackage{bm}

\usepackage{color}
\usepackage{xcolor}
\usepackage{ctable}
\usepackage[printwatermark]{xwatermark}

\newcommand{\heading}[1]{\vspace{0.25truecm}\noindent\textbf{#1}}
\newcommand{\quot}[1]{``#1''}

\usepackage{colortbl} 
\usepackage{booktabs,siunitx}
\newcommand{\quantities}[1]{%
  \begin{tabular}{@{}c@{}}\strut#1\strut\end{tabular}%
}

\definecolor{NavyBlue}{RGB}{38,94,196}
\definecolor{Crimson}{RGB}{211,31,45}

\begin{document}

\title{Influence of augmented humans in online interactions during voting events}


\author{Massimo Stella}
\affiliation{Fondazione Bruno Kessler, Via Sommarive 18, 38123 Povo (TN), Italy}
\author{Marco Cristoforetti}
\affiliation{Fondazione Bruno Kessler, Via Sommarive 18, 38123 Povo (TN), Italy}
\author{Manlio De Domenico}
\email[Corresponding author:~]{mdedomenico@fbk.eu}%
\affiliation{Fondazione Bruno Kessler, Via Sommarive 18, 38123 Povo (TN), Italy}

\begin{abstract}
The advent of the digital era provided a fertile ground for the development of virtual societies, complex systems influencing real-world dynamics. Understanding online human behavior and its relevance beyond the digital boundaries is still an open challenge. Here we show that online social interactions during a massive voting event can be used to build an accurate map of real-world political parties and electoral ranks. We provide evidence that information flow and collective attention are often driven by a special class of highly influential users, that we name ``augmented humans'', who exploit thousands of automated agents, also known as bots, for enhancing their online influence. We show that augmented humans generate deep information cascades, to the same extent of news media and other broadcasters, while they uniformly infiltrate across the full range of identified groups. Digital augmentation represents the cyber-physical counterpart of the human desire to acquire power within social systems.
\end{abstract}

\maketitle

\section*{Introduction}

Online social actions drive collective attention and dynamics~\cite{de2013anatomy,borge2016dynamics}, having a deep impact on the construction and perception of social reality. Many large-scale studies have reported evidence of online ecosystems altering decision-making of crowds~\cite{muchnik2013social} and influencing real-world voting of millions of people~\cite{bond201261}. The last few years have seen a deluge of increasingly more sophisticated automated online agents, called also ``bots'', populating techno-social systems cleverly disguised as human users~\cite{wagner2012social,cresci2015fame,cresci2017paradigm,gilani2017classification,varol2017online}. Nowadays, bots can produce credible content with human-like temporal patterns~\cite{ferrara2016rise,ferrara2017disinformation,stella2018bots}. By promoting online activity, bots can interact with humans and influence their standing against specific topics such as political issues~\cite{ferrara2016rise,cresci2017paradigm,stella2018bots,vosoughi2018spread}. Since manoeuvring social platforms can deeply affect real-world dynamics~\cite{gonzalez2011dynamics,gonzalez2013broadcasters}, understanding if and how computer-generated activities can alter the behavioral responses of humans to achieve online social manipulation is of utmost importance~\cite{aral2009distinguishing,wagner2012social,aral2012identifying}. Identifying and quantifying these effects is particularly crucial during voting events, where individuals' decisions might be driven by external events, such as natural disasters or economic shocks~\cite{ashworth2018learning}. While attention is generally paid to how physical interactions among voters and electoral arrangements influence voting behavior, Bruter and Harrison~\cite{bruter2017understanding} shifted the focus on the psychological influence that electoral arrangements exert on voters by altering human emotions and behavior. The investigation of voting from a cognitive perspective leads to the concept of \textit{electoral ergonomics}: Understanding the optimal ways in which voters emotionally cope with voting outcomes can lead to a better prediction of the elections.

Here we quantify to which extent online social activity reflects the real world and we characterize the peculiar behavior of a class of individuals who make a massive use of bots to enhance their online visibility and influence. The term \textit{cyborg} has been used in this context to identify, indistinctly, bot-assisted human or human-assisted bot accounts generating spam content over social platforms such as Twitter~\cite{wagner2012social,chu2010tweeting}. Here, we prefer to use the term \textit{augmented human} for indicating specifically those human accounts exploiting bots for artificially increasing, i.e.  \textit{augmenting}, their influence in the digital world, analogously to physical augmentation improving human performances in the real world~\cite{savulescu2009human}. 
Like several automated agents identified in our data set, augmented humans played a special role for information spreading, by triggering deep information cascades with the help of bots.

\section*{Results}

We collected 966,483 messages posted to the microblogging platform Twitter from 194,273 unique users during Italian elections (see Methods for details). The data was collected from February 24 2018 until March 5 2018.

\subsection*{Bot identification}

To identify automated agents in the data set, we developed a deep neural network model (see Methods and SI), which classified 13.4\% of users as bots, a value compatible to estimations during other voting events~\cite{ferrara2016rise,ferrara2017disinformation,stella2018bots}. We built the network of interactions between human users and bots, including different types of social actions such as Retweets (i.e. a user sharing another user's message), Mentions (i.e. a user mentioning another use in a message) and Replies (i.e. a user starting a discussion with another user). While Mentions and Replies can have both negative and positive connotations, Retweets are traditionally considered as a form of social endorsement \cite{metaxas2015retweets,aral2012identifying}: Users tend to retweet and thus endorse content they agree with. 

\subsection*{Human-bot interactions: Homophily and centrality}

\begin{figure*}
    \centering
        \includegraphics[width=0.95\textwidth]{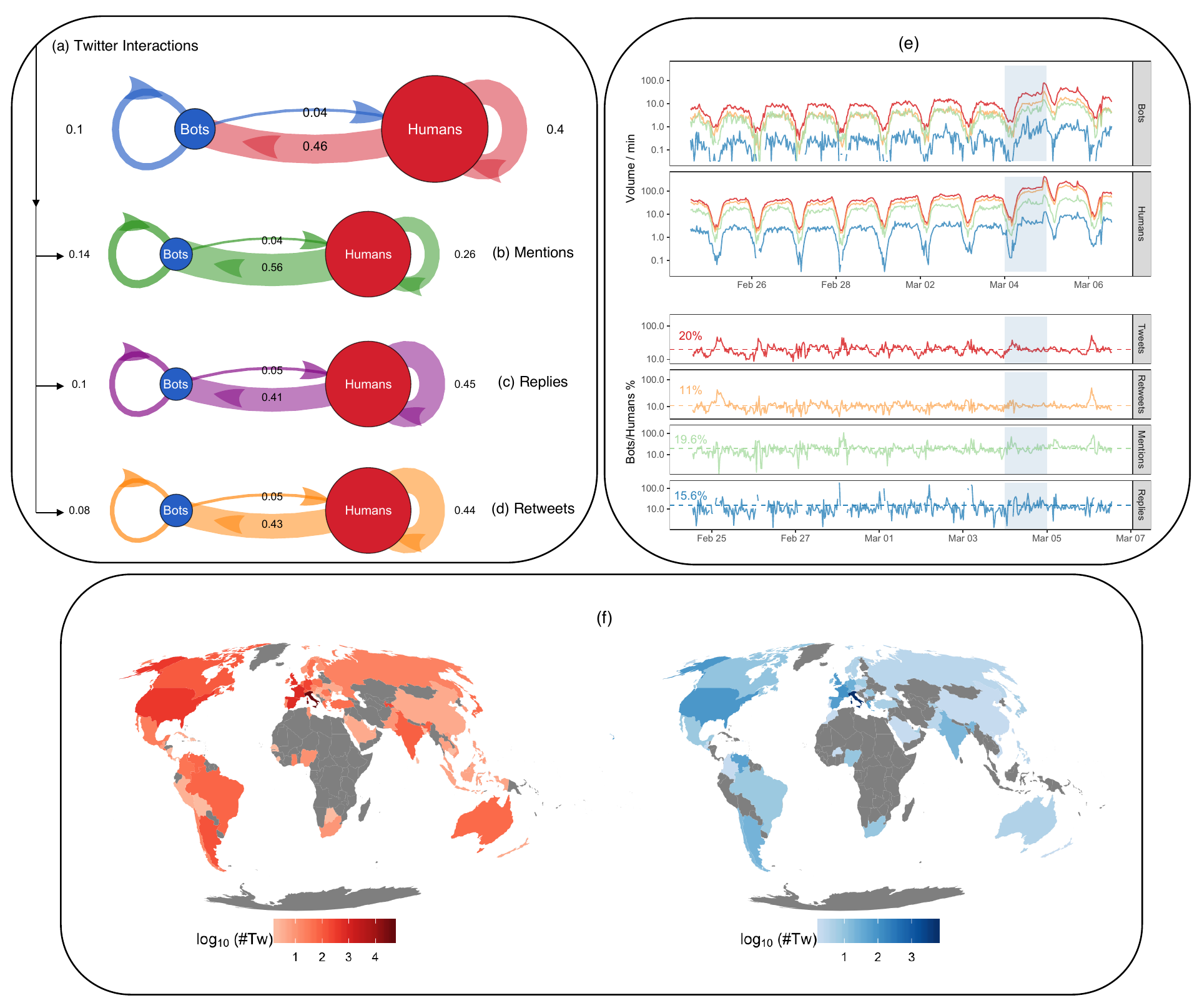}
    \caption{\label{fig:1}\textbf{Online human-bot interactions during the Italian elections.} \textbf{(a)}: Volumes of human-bot interactions in Twitter. \textbf{(b-d)}: Human-bot interactions stratified by actions: Mentions, Replies and Retweets. \textbf{(e)}: Evolution across time of the overall social activity of humans and bots (top), also stratified by actions (bottom). \textbf{(f)}: Geographic location of involved users, where the color encodes the number of tweets per country, in logarithmic scale. As in (a), humans are in red and bots are in blue. Users are mostly located in Italy, with relevant interactions from other countries worldwide.}
\end{figure*}

Figure~\ref{fig:1} shows the temporal evolution of volumes for messages (i.e., Tweets) and the considered social actions for both bots and humans. Figure~\ref{fig:1} (a-d) indicates the overall fraction of messages exchanged between bots and humans (a), and the fractions stratified by social interactions (b-d). Most of the social interactions are from humans to bots (46\%); Humans tend to interact with bots in 56\% of mentions, 41\% of replies and 43\% of retweets. Bots interact with humans roughly in 4\% of the interactions, independently on interaction type. This indicates that bots play a passive role in the network but are rather highly targeted by humans.
Figure~\ref{fig:1} (e) shows the number of social interactions over time. The circadian rhythm is evident, i.e. at night the volume of messages generated by humans drops down. Also bots display a similar circadian rhythm, in agreement with previous observations~\cite{ferrara2016rise,stella2018bots}. In general, bots contribute to 6\% of the total number of social interactions occurred during the voting event (March 4 2018).
Figure~\ref{fig:1} (f) reports the geographic locations of both human and bot users in the social system. Although most of the users are located in Italy, there are significant fractions of human users also located in the United States and in Europe, indicating the worldwide relevance of the Italian voting. Similarly, bots' locations are distributed worldwide and they are present in areas where no human users are geo-localized such as Morocco, Peru, Finland or Indonesia.  

The analysis of observed social interactions (links) between users (nodes) before, during and after the voting day revealed bot homophily, i.e., automated agents tend to interact more with other bots rather than with humans compared to random expectation  (see SI). Since interactions encode content spread~\cite{aral2009distinguishing}, this result indicates that bots share messages mainly with each other and hence can resonate with the same content, be it news or spam. Furthermore, if we quantify the centrality of a user in terms of the probability of finding it by exploring the web of interactions at random, then we find that bots are almost twice as central as humans (see SI). The above findings indicate that bots play the role of sinks for information flow. In fact, 9 out of 10 hubs -- i.e., highly interacting users -- are bots and they are mainly news media and public profiles of politicians, which usually act as broadcasters and drive online information flow~\cite{aral2012identifying,gonzalez2013broadcasters}.  The analysis of topic frequency and associations in bot-generated messages confirms this trend: Bots act as broadcasters by repeating the same political content of human users, boosting the spread of hashtags related to the electoral process (see SI).

\subsection*{Information cascades identify different classes of influencers}

The observed social interactions build a complex network with a heterogeneous connectivity distribution. Such systems are well known for being susceptible to cascading events~\cite{watts2002simple,gleeson2017temporal} and, in the case of online social networks, the phenomenon might manifest as collective action and faster diffusion of specific information \cite{goel2012structure,de2013anatomy,borge2016dynamics,martin2016exploring}. In information cascades, a single piece of information is originated by a seed user, it is endorsed by other users in his/her neighborhood and consequently re-shared across the network~\cite{goel2012structure}. Cascade size depends on a variety of factors~\cite{gleeson2016effects}, including -- but not limited -- to the structure of the network and the information content, making their prediction rather difficult~\cite{martin2016exploring}. 

We have tracked 83,593 information cascades during Italian elections and, for each one, we have analyzed the underlying structure by measuring its size, i.e. the number of times an information has been re-shared, and its diameter, i.e. the maximum topological distance, not accounting for the directionality of the dynamical process. As expected for complex networks with highly heterogeneous connectivity~\cite{watts2002simple}, the distribution of observed cascade sizes is heavy-tailed and compatible with a power law characterized by a scaling exponent $\gamma=-2.33 \pm 0.04$, similarly to size distribution in percolation theory or avalanches in self-organized criticality~\cite{dorogov2008}. Cascade size ranges between 2 and 4,313.

\begin{figure*}[ht]
\centering
\includegraphics[width=\textwidth]{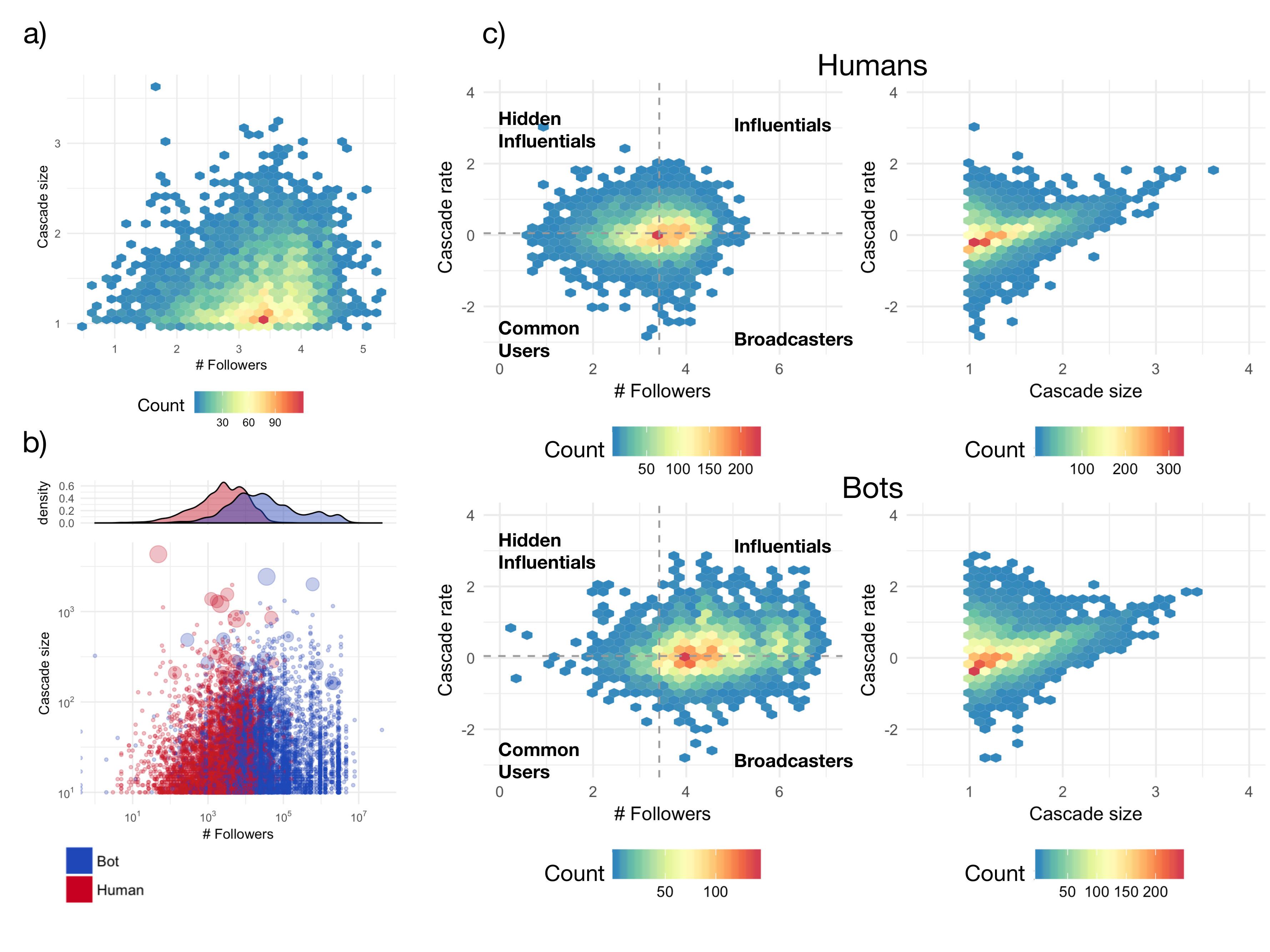}
\caption{\textbf{Information cascades during Italian elections.} a) Heatmap of the number of users initiating information cascades, as a function of the size of their social neighborhood (Followers) and the size of the generated cascade; b) Scatter plot of the same data, with points encoding users. Color encodes bot/human classification and size encodes cascade's diameter; c) As in a) but considering cascade rate, defined by the ration between cascade size and its duration, vs. neighborhood size (left panels) and cascade size (right panels), for humans (top panels) and bots (bottom panels). The heatmap of cascade rate vs. neighborhood size allows one to identify 4 categories: hidden influentials, influentials, common users and broadcasters (see the text for further detail). Dashed lines indicate medians of structural and dynamical features in humans. Only cascades with at least 10 adopters are considered and, for heatmaps, the logarithm of the corresponding variables is considered.}
\label{fig:cascades-hexmap}
\end{figure*}

We show in Fig.~\ref{fig:cascades-hexmap} (a) a heat map of cascade size vs. the size of initiators' social neighborhood (i.e., the number of followers). As expected, on average, larger the number of followers larger the cascade size, with very few exceptions. Figure~\ref{fig:cascades-hexmap} (b) shows the same data, with explicit information about user classification and the cascade diameter, ranging between 2 and 6. This figure shows a good separation between human and bot behavior. Deeper information cascades are generated mostly by humans with a high number of followers, with the remarkable example of one, User01, who produced the largest cascade among humans and bots despite having less than 100 followers.

Recently, dynamical activity-connectivity maps based on network and temporal activity patterns -- or their variation -- have been used to identify influential individuals or broadcasters during online protest diffusion~\cite{gonzalez2013broadcasters} and contagion dynamics of extremist propaganda~\cite{ferrara2017contagion}. For instance, Bastos and Mercea~\cite{bastos2016serial} used hashtag trends for showing the existence of ``serial activists'', users with ordinary numbers of followers but very prolific in producing content about multiple political topics and bridging together disparate communities. Gonzalez et al. \cite{gonzalez2013broadcasters} related topological properties, such as the ratio between incoming (friends) and outgoing (followers) connections, to dynamical properties, such as the ratio of received and posted messages.

Here, we argue that it is also plausible to relate individuals' social influence to the size of information cascades they generates with their content~\cite{bakshy2011everyone}. To this aim, we propose a more complex map relating a topological feature, i.e. the number of outgoing connections (followers), and a dynamical feature, i.e. information cascade growth rate, defined by the ratio between a cascade size and its duration over time. Baseline social behavior during a specific event, such as the Italian election in our case, is defined by the medians of the two observables, like shown in Fig.~\ref{fig:cascades-hexmap} (c). This map allows to easily identify four categories of individuals in the social dynamics: i) \textit{hidden influentials}, generating information cascades rapidly spreading from a small number of followers; ii) \textit{influentials}, generating information cascades rapidly spreading from a large number of followers; iii) \textit{broadcasters}, generating information cascades slowly spreading from a large number of followers; iv) \textit{common users}, generating information cascades slowly spreading from a small number of followers. Remarkably, the topological and dynamical behavior of humans and bots is very different: during Italian elections, bots are mostly broadcasters (mostly media) and influentials (mostly political leaders). Figure \ref{fig:cascades-hexmap} (c) (left) highlights a positive correlation between cascade rate and size: Cascades involving more users tend also to flow over the interactions web at faster rates. This positive trend is stronger for cascades of sizes larger than $10^2$. The stronger correlation for larger cascades suggests that they differ qualitatively from smaller cascades: Larger cascades contain specific semantic content, in this case political-related topics, which accelerate spreading.

\subsection*{The social bulk of endorsements mirrors political antagonism}

So far, our analysis characterized online human behavior in terms of human-bot interactions and information spreading. However, to quantify to which extent the observed online social activity reflects the real world a more sophisticated analysis is required. To this aim, we analyzed the static representation of the system, where interactions across time have been aggregated to a directed and weighted social network. We then identified the core of the observed social system by tracking the most relevant interactions among the most important actors. We identified relevant interactions by assuming that if two users share similar political ideologies, they can endorse and subsequently share (i.e. retweet) the content of each other. However, if only re-sharing was considered, the network would contain a lot of spurious connections due, for instance, to fortuitous endorsement rather than to a systematic intention. 

We first filtered the network by considering only pair of users with at least one retweet, with either direction, because re-sharing content it is often a good proxy of social endorsement~\cite{metaxas2015retweets}. We then considered a more selective restriction, by requiring that at least another social action -- i.e., either mention or reply -- must be present in addition to a retweet. This restrictive selection allows one to filter out all spurious interactions among users with the advantage of not requiring any threshold with respect to the frequency of interactions themselves. The resulting network is what we call the \emph{social bulk}, i.e. a network core of endorsement and exchange among users. By construction, information flows among users who share strong social relationships and are characterized by similar ideologies: In fact, when a retweet goes from one user to another one, both of them are endorsing the same content, thus making non-directionality a viable approach for representing the endorsement related to content sharing. Therefore, in the following, we can safely consider undirected interactions among users. Connections between users are weighted by the aggregated frequency of their social interactions. An illustration of how the social bulk is built is shown in Fig.~\ref{fig:segreg-golden} (a).

In the following, we introduce different measures to quantify different features of the social bulk, i.e. social polarization, fragmentation and segregation. 

The concept of \textit{social polarization} assumes the existence of two competing stances or opposing groups characterizing the mesoscale organization of the system~\cite{conover2011political}. In presence of two groups, they can be identified by calculating, for instance, the Fiedler partitioning~\cite{ding2001min,stella2018bots}, which is related to the min-max cut problem for finding optimal flows in networks~\cite{ding2001min}. Fiedler partitioning (see Methods) separates the users of a connected graph into two classes such that the total number of inter-class connections is close to the optimal minimum. If interactions encode strong social relationships, as in the social bulk, then the Fiedler partitioning identifies two factions antagonizing each other by sharing the least endorsements possible. 

We measure social polarization by computing the modularity~\cite{newman2006modularity} of the social bulk with respect to its Fiedler partitioning (see Methods). The larger the modularity $\Phi$ of the Fiedler partitioning, i.e. system polarization, the more antagonized are users into two opposing groups. For the largest connected component of the social bulk we calculate the polarization $\Phi_{F}=0.452$. The expected polarization of a null model -- where social relationships are uniformly randomized while preserving the individual degrees and the distribution of strengths -- is $\left\langle \Phi_{F}^{rand} \right\rangle = 0.301$, significantly different from the observed network (p-value~$<10^{-5}$). This result indicates that the heterogeneity of social interactions can not explain, alone, the observed level of polarization, which has rather to be attributed to other causes such as political parties or opposing political trends. 

However, during the Italian elections considered in this work more than two political parties were present, so that the notion of polarization has to be extended to account for the presence of several opposing groups. For complex networks, a widely adopted approach is to use modularity maximization for group identification~\cite{newman2006modularity,fortunato2010community,fortunato2016community,newman2012communities}. Identified communities of users are characterized by intra-group connectivity denser than inter-group one. 

\begin{figure}[!t]
\centering
\includegraphics[width=0.47\textwidth]{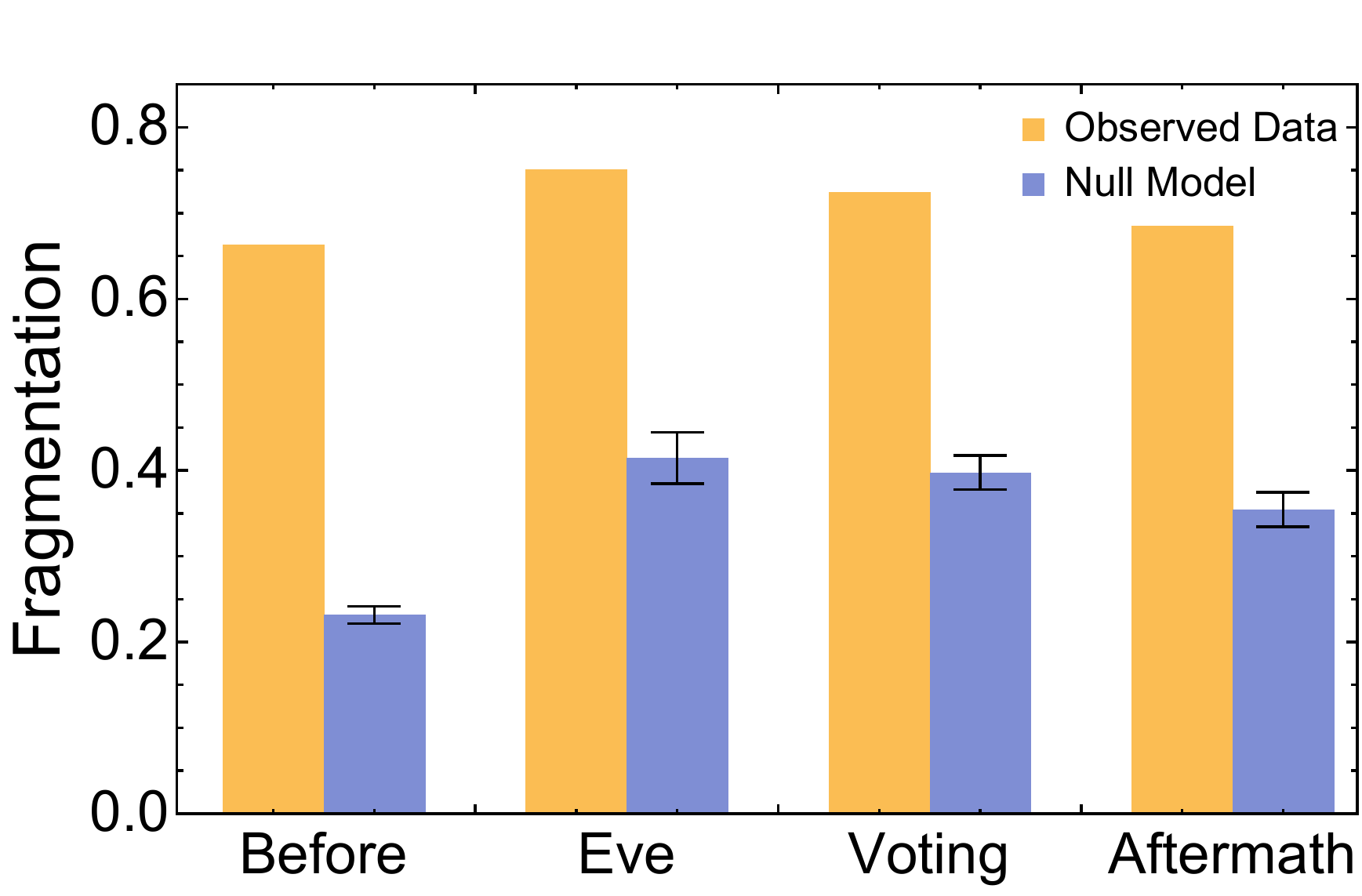}
\caption{\textbf{The online system is characterized by social fragmentation.} \textbf{Top}: Fragmentation encodes the tendency of online users to organize in multiple opposing groups (see the text for further detail). During the four considered periods, the online social network is fragmented much more than random expectation. Small changes in fragmentation of the observed system across time are reflected in the null model, indicating that they can be explained by small changes in the heterogeneity of the underlying connectivity.}
\label{fig:segreg}
\end{figure}

In the case of the social bulk we can interpret modularity as an estimation of system's \textit{social fragmentation} into more than two opposing groups. Here, we use the Louvain multilevel approach, known to be very efficient on large-scale networks~\cite{blondel2008fast}. In the whole social bulk we measure a fragmentation $\Phi_{L}=0.812$, indicating the presence of several factions in the bulk network that are in a stronger opposition when compared to the null model ($\left\langle \Phi_{L}^{rand} \right\rangle = 0.692$, p-value~$<10^{-5}$). The fact that $\Phi_{L}>\Phi_{F}$ indicates that a more accurate description of the mesoscale organization of the social bulk is given when more than two groups are considered, in agreement with our hypothesis that results should reflect the real world socio-political scenario. To understand if this finding is robust or just an artefact due to how the social bulk is built, we have measured the social fragmentation of the original system during all phases of voting (see Fig.~\ref{fig:segreg}). Once again, we observe that social fragmentation is stable across time and significantly larger than random expectation, confirming that results obtained from the social bulk are consistent.

However, neither polarization nor fragmentation can be used to quantify to which extent the system consists of isolated groups -- the ones with no interactions with the rest of the system -- which are effectively segregated in the network. Note that we are not referring to users in the periphery~\cite{borgatti2000models,rombach2017core} of the system, where information can slowly but flawlessly flow among all nodes in the network. Instead, we refer to groups unable to exchange information with the core of the system, i.e., to nodes belonging to disconnected components. We quantify \textit{social segregation} $\Sigma$ by considering the average number of connected components weighted by the number of their links (see Methods). If a social network consists of isolated nodes only, then $\Sigma=1$, whereas $\Sigma=0$ for systems with a single connected component. For a network consisting of $M$ connected components of same size and density of interactions $\Sigma=1 - 1/M$: The larger the number of components, the larger the social segregation. The segregation of the social bulk is $\Sigma=0.476$, significantly stronger than random expectation $\langle\Sigma^{rand}\rangle=0.172$ (p-value~$<10^{-5}$) based on a configuration model preserving the connectivity distribution. This indicates that strong interactions lead to more segregated components, with fewer bridges among connected components than expected from the heterogeneity of interactions only. Hence, the observed segregation represents additional evidence for the presence of antagonism in the considered social ecosystem.

Polarization, fragmentation and segregation analyses all constitute evidence that the social bulk displays densely connected groups in opposition with each other. 

\subsection*{Groups in the social bulk highlight digital augmentation}

Through the multilevel approach, we identified 8 main opposing communities (i.e. having more than 2\% of the total nodes in the network), as reported in Tab.~\ref{tab:commbulk}. The analysis of hubs in each group of the social bulk indicates that i) one group corresponds to a single augmented human and his/her bots; ii) five groups directly map the ecosystems of the main Italian political parties; iii) two groups encode news media universe, either traditional or online news organisations. 

In this context, we provide an operative definition of augmented humans as human users having at least 50\% of bot neighbours in the social bulk. Users with less than 3 bulk interactions are discarded. We systematically identified 1,010 user accounts ($12.7\%$ of humans in the social bulk) corresponding to augmented humans. The most central augmented human in terms of number of social interactions is User01 which interacts with 2,700 bots and 55 humans in the social bulk. We have anonymized the username for privacy purposes.

It is natural to wonder about how bots, humans and augmented humans are organized into communities within the social bulk. In fact, given the relevance of the voting event in the real world, our hypothesis is that communities should reflect real political movements and groups, to some extent. 

\begin{figure*}[ht]
\centering
\includegraphics[width=16cm]{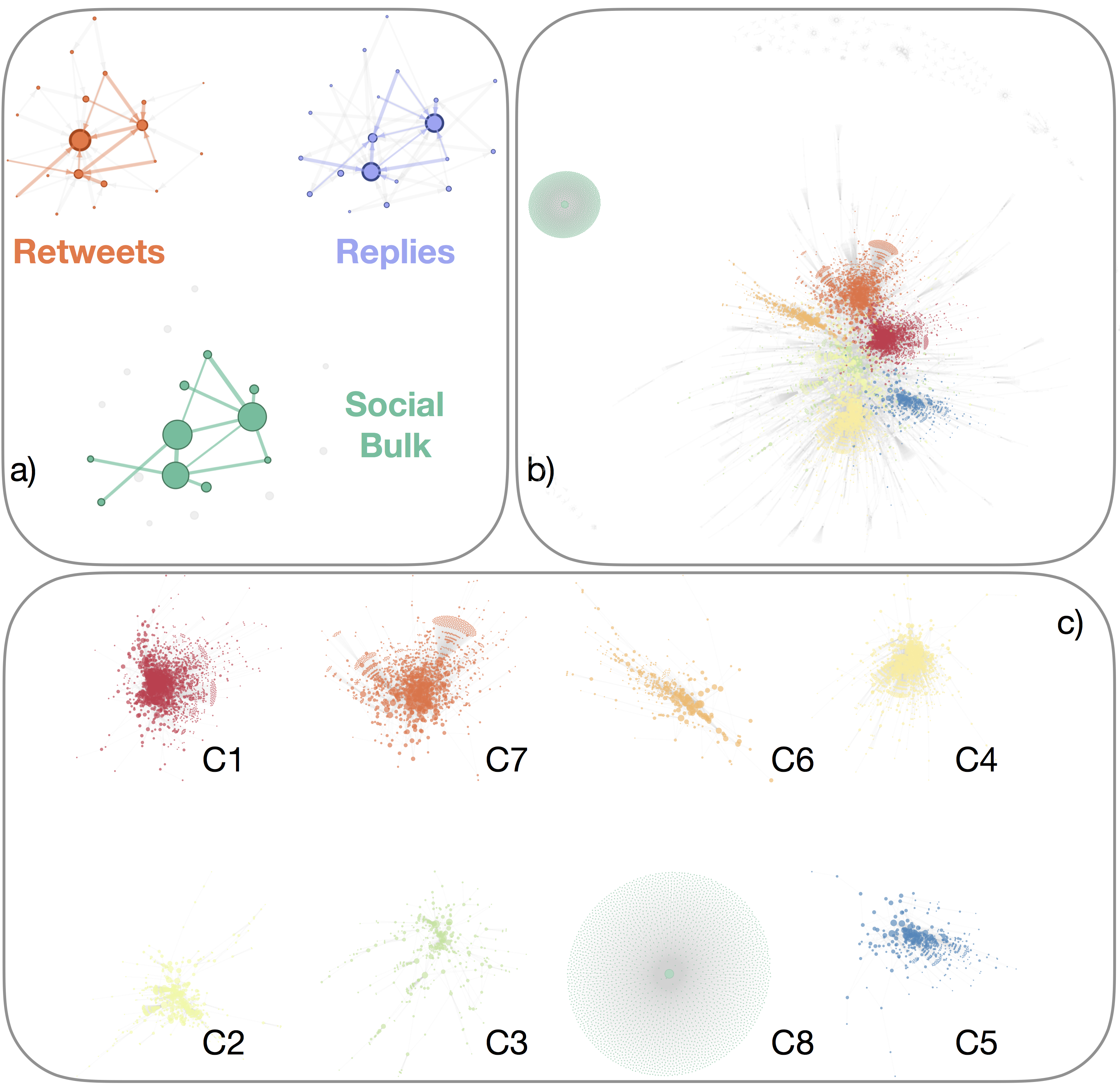}
\caption{\textbf{Social bulk of Italian elections.} \textbf{a)} Twitter users can retweet or mention or reply with each other. Each action encode a specific social meaning and, by considering the co-existence of endorsement (i.e. retweet) and discussion (i.e. mention or reply), between the same pairs of users, we filter out spurious interactions to identify the social bulk of the system. \textbf{b)} Visualization of the social bulk emerged during Italian elections, with users (i.e., the nodes) colored by the community they belong to (see the text for further detail). \textbf{c)} The eight communities with at least 2\% of users are represented separately, while preserving their relative position in the social bulk shown in panel b). Note the remarkable star-like topology characterizing the augmented human identified in the system.}
\label{fig:segreg-golden}
\end{figure*}

First, we focused our attention on the augmented human's group, consisting of more than 2,500 automated agents artificially interacting with the augmented human user. This peculiar activity leads to a star-like structure for the corresponding community, as shown in Fig,~\ref{fig:segreg-golden}. This finding has triggered our attention, driving our efforts towards quantifying the infiltration $\mathcal{I}_{s}$ of a specific class $s$ of accounts in each group, by considering the corresponding fraction of users in a given group (see Methods). Table \ref{tab:parties} reports the infiltration of augmented humans in the groups of the social bulk. Unsurprisingly, infiltration of bots is higher in the group representing the augmented human and his/her automatic entourage of interacting social bots. Furthermore, we find that groups relative to news media are richer in bots compared to groups representing political parties, which is compatible with our previous finding of bots being preferentially news media broadcaster in the observed data. 

The infiltration of augmented humans is approximately uniformly distributed across all identified groups, with the remarkable exception of C8, the augmented human's community. One would expect for the groups richer in bots to have also more augmented humans. Instead, bot and augmented human infiltration do not correlate with each other (Kendall Tau 0.07, p-value $0.8$), indicating that augmented humans tend to interact selectively with the bots available in their groups rather than creating more. This trend is not valid for the group C8, where one human (User01) interacts almost exclusively with bots. 

\subsection*{Augmented humans are hidden influencers}

All the augmented humans identified in this study have, on average, less than 9,000 followers and 1,500 friends, indicating that a considerable amount of social influence was obtained by users that preferentially interacted with bots during the considered event. The analysis of information cascades revealed that almost 2 out 3 augmented humans played an important role in the flow of online content: 67\% of this class of users were either influentials or hidden influentials or broadcasters. Hidden influentials, known to be efficient spreaders in viral phenomena~\cite{banos2013role}, are mostly humans but augmented humans also falls in this category (e.g. User01). 

\begin{table*}
\begin{centering}
\begin{tabular}{|c|c|c|c|c|}
\hline 
Bulk Group & Top Influencers & Social Ecosystem & \quantities{Bot \\ infiltration} & \quantities{Augmented \\ infiltration}\tabularnewline
\hline 
\hline 
C1 & \quantities{fattoquotidiano, User02, Mov5Stelle, MovPopulistaIta, \\ M5S\_Europa,
Puglia\_M5S, ManlioDS, MovPopulistaIta} & M5S & 13.4\% & 13.3\% \tabularnewline
\hline 
C2 & \quantities{repubblica, Agenzia\_Ansa, sole24ore, TgLa7, RaiNews, \\ Viminale, RTL1025,
Corriere, MediasetTgcom24} & Media & 24.5\% & 12.1\% \tabularnewline
\hline 
C3 & \quantities{SkyTG24, you\_trend, nonleggerlo, ilpost, \\ claudioporcu73, User03,
agorarai, espressoline} & Web Media & 26.6\% & 15\% \tabularnewline
\hline 
C4 & \quantities{matteorenzi, pdnetwork, CarloCalenda, EugenioCardi,\\ danielevpd, PaoloGentiloni, PdMilano} & PD & 15.5\% & 15.4\% \tabularnewline
\hline 
C5 & \quantities{liberi\_uguali, civati, PietroGrasso, lauraboldrini,\\ LiberiEugualiIT,
SI\_sinistra} & LEU & 16.3\% & 19.1\% \tabularnewline
\hline 
C6 & \quantities{berlusconi, forza\_italia, Elezioni2018\_FI,\\ renatobrunetta, MennaFini,
GruppoFICamera} & FI & 28.2\% & 22.1\% \tabularnewline
\hline 
C7 & \quantities{RiscattoNaz, mattosalvinimi, borghi\_claudio, AlbertoBagnai,\\GiorgiaMeloni,
FratellidItalia, LegaSalvini} & Lega and FdI & 9.5\% & 12.9\% \tabularnewline
\hline 
C8 & User01 & Augmented Human & 97.9\% & $3\cdot 10^{-4}$ \% \tabularnewline
\hline 
\end{tabular}
\par\end{centering}

\protect\caption{\label{tab:commbulk}\textbf{Largest online social groups.} Most populated communities (with more than 250 users) in the social bulk, with top influencers listed per group. Top influencers are identified as hubs in the bulk network. As evident from the similarities among top influencers, groups reflect specific ecosystems of the Italian voting event: ``Movimento 5 Stelle'' (M5S), traditional media (Media), media with massive online presence (Web Media), ``Partito Democratico'' (PD), ``Liberi e Uguali'' (LEU), ``Forza Italia'' (FI), ``Lega'' and ``Fratelli d'Italia'' (Lega and FdI), and then the augmented human with all his/her interacting bots (Augmented Human). Bot (augmented) infiltration indicates the percentage of bot users (augmented humans) in each group. Excluding the community corresponding to the augmented human (made for 97.9\% of bots), the mean bot infiltration in the bulk network is $29.2\%$ while the mean augmented infiltration is 15.7\%. The media groups are richer in bots as expected, since they include news media and online accounts of news papers.}
\end{table*}

\subsection*{Groups in the social bulk reflect electoral outcomes} 

In order to investigate the representativeness of online groups in terms of real-world events beyond the hub analysis, we focused on the structural features of groups, namely the interaction volume of a group (i.e. the number of strong social interactions among users in the group) and the group size (i.e. the number of users in a given group). In Tab.~\ref{tab:parties}, we show that the outcome of Italian elections (i.e. the fraction of votes received by each political group) strongly correlates with the group volume (Spearman rank correlation coefficient $\rho=0.9$, p-value~$=0.039$). This correlation is statistically significant within a 5\% significance level and direct sampling of rankings was used in order to compute the p-value without relying on any assumption about the large-scale statistical properties of the data. The strong correlation found indicates that the volume of online interactions closely mirrors the election outcome. 

\section*{Discussion}

Online social systems and the information they continuously generate provide an invaluable resource for computational social scientists and their large-scale analysis of human behavior~\cite{lazer2009life,conover2011political,ruths2014social} and the emergence of collective attention~\cite{borge2016dynamics,baronchelli2018emergence}. The analysis of information and behavioral spreading on social media~\cite{aral2012identifying} revealed that an individual is much more likely to adopt a content when his/her neighbors in the social network tend to reinforce it~\cite{centola2010spread}. On the one hand, this allows online media to facilitate, for instance, the dissemination of emergency information and help coordinate relief efforts~\cite{kryvasheyeu2016rapid}. On the other hand, the same social networks can be misused to spread fake content farther, faster and deeper~\cite{vosoughi2018spread}.

In this work we have identified and quantified a new phenomenon, i.e. digital augmentation, to characterize individuals that coordinate from hundreds to thousands of social bots for achieving a social influence comparable to the one of political parties and news media organisations, with serious repercussions in the real-world. 

Our results strongly support the idea that via augmentation even common users can become social influencers without having a large social neighborhood but rather by recurring to the aid of either armies of bots or the selection of a few key helping bots. This digital augmentation represents an interesting behavioral response aimed at overcoming the well documented pressure for achieving influence and recognition in online ecosystems \cite{aral2009distinguishing,cresci2015fame,bond201261,gleeson2016effects} and during voting events \cite{bruter2017understanding}. While in real life such augmentation comes mainly from smart devices, our work presents compelling evidence that in online social platforms the augmentation for achieving social influence is represented by an exploitation of social bots by human accounts. 

Furthermore, the strong correlation between the volume of online interactions in the social bulks and the electoral outcomes highlights the role potentially played by online social systems during the voting process. This finding is in full agreement with previous works showing how online ecosystems acted upon society by altering the emotions \cite{bruter2017understanding} and beliefs \cite{muchnik2013social,bond201261} of large populations of individuals. It is worth underlining that the observed groups are relative to the network structure of social endorsements: Considering the layout of online endorsement can provide information beneficial for more accurate predictions of electoral outcomes. Further investigation of online social systems under the perspective of predicting electoral outcomes would provide interesting challenges for future work.

Our work provides a first step towards a more systematic quantification of the impact of digital augmentation in opinion formation and the manipulation of online attention by means of human-bot interactions.

\begin{table*}
\begin{centering}
\begin{tabular}{|c|c|c|c|c|c|c|}
\hline 
Group & Election Outcome (\% votes) & Rank & Interaction Volume & Rank & Group Size &  Rank\tabularnewline
\hline 
\hline 
M5S & 32.68\% & 1 & 3,857 (34.82\%) & 1 & 1,162 & 2\tabularnewline
Lega and FdI & 20.76\% & 2 & 3,205 (28.93\%) & 2 & 1,491 & 1\tabularnewline
PD & 18.72\% & 3 & 2,504 (22,60\%) & 3 & 1,133 & 3\tabularnewline
FI & 14.01\% & 4 & 675 (6.09\%) & 5 & 330 & 5\tabularnewline
LEU & 3.39\% & 5 & 837 (7.55\%) & 4 & 434 & 4\tabularnewline
\hline 
\end{tabular}
\par\end{centering}
\protect\caption{\textbf{Network analysis of groups in the social bulk reflect election outcomes.}
The five political ecosystems from the bulk network are ranked against their topological features: i) interaction volume, i.e. the number of social actions within the group; ii) size, i.e. the number of individuals in the group. The rank based on online interactions strongly mirrors the election outcome (Spearman $\rho=0.9$, p-value~$=0.039$), supporting the hypothesis that online social interactions are tightly entwined to outcomes and events in the real-world. \label{tab:parties}
}
\end{table*}

\section*{Methods}

\heading{Data collection.} Between 24 February 2018 and 7 March 2018, we have collected 966,483 messages (tweets) posted by 194,273 different users to the microblogging platform Twitter, containing at least one of the following keywords or hashtags: \textit{\quot{elezioni}, \quot{\#elezioni}, \quot{\#elezioni2018}, \quot{\#elezioni4marzo}, \quot{\#ItalyElection2018}, \quot{\#voto}, \quot{\#4marzo}, \quot{\#M5S}, \quot{\#PD}, \quot{\#LeU}, \quot{\#LiberieUguali}, \quot{\#ForzaItalia}, \quot{\#FDI}, \quot{\#FI}, \quot{\#lega}, \quot{\#FratellidItalia}, \quot{\#MDP}}.

Tweets have been collected using the streaming real-time provided by Twitter API platform, filtered by the above keywords. Twitter by default limits to 1\% of the overall number of Tweets per second the fraction of tweets that can be retrieved from the streaming API. However, when the fraction of tweets concerning specific keywords is smaller than 1\% of the global volume, Twitter does not apply limitations and the complete flow of information is collected. When this is not the case, Twitter provides messages of warning, reporting the cumulative number of missed tweets.

In the case of Italian elections, we received no warnings, therefore we have collected 100\% of tweets containing the specified keywords.

\heading{Classification task.} In this work the classification of users in our data set as \quot{humans} or \quot{bots} is based on features providing the best classification accuracy according to recent studies~\cite{ferrara2017disinformation}: 1) \textit{Statuses count}; 2) \textit{Followers count}; 3) \textit{Friends count}; 4) \textit{Favourites count}; 5) \textit{Listed count}; 6) \textit{Default profile}; 7) \textit{Geo enabled}; 8) \textit{Profile use background image}; 9) \textit{Protected}; 10) \textit{Verified} for a total of ten features ($N_{feats}=10$).

Searching for better performance we tested different machine learning techniques on an independent dataset created ad-hoc (see Supplementary Information) from a collection of manually annotated datasets (see Tab~\ref{tab:dataset_train}). Models are trained on the 80\% of the data and validated over the remaining 20\%. The subdivision between the two sets was carried respecting the balancing between bots and humans present at the level of the single original datasets, in this way we have all type of different bots both in training and validation. The models based on random forest and deep neural network provided us with the highest accuracy ($> 90\%$) and precision in identifying bots ($> 95\%$). We chose the deep neural network model because it also provided a more stable classification of certain users playing the role of broadcasters (see Supplementary Information).

\begin{table}
\begin{centering}
\begin{tabular}{|c|c|c|c|}
\hline 
data set & bot & human & total\tabularnewline
\hline 
\hline 
cresci2015     &     0 &  5301 &   5301 \tabularnewline
cresci2017     &  7543 &  3474 &  11017 \tabularnewline
cyborgs        &  2756 &     0 &   2756 \tabularnewline
aboutme        &     0 &  2463 &   2463 \tabularnewline
omnibots       &  3530 &     0 &   3530  \tabularnewline
russian-trolls &   389 &     0 &    389 \tabularnewline\hline
\hline 
TOTAL & 14218 & 8775 & 22993\tabularnewline
\hline 

\end{tabular}
\par\end{centering}
\protect\caption{\textbf{Proportions of bot and human users in the training data.}
}
\label{tab:dataset_train}
\end{table}

\heading{Fiedler partitioning, modularity, segregation and infiltration.} Fiedler partitioning is a widely used technique from spectral graph theory for solving the min-max cut problem, i.e. partitioning a network in two components of similar size but connected by links whose total weights are the smallest possible \cite{ding2001min}. Fiedler partitioning is obtained by considering the eigenvalue problem:

\begin{equation}
(D-W)\mathbf{q}=\lambda \mathbf{q},
\end{equation}

for a connected network represented by the weighted adjacency matrix $W$, with $w_{ij}$ equal to the weight of the link between nodes $i$ and $j$, and by a matrix $D$ having the strength of nodes on its main diagonal. The spectral partitioning is obtained by identifying nodes relative to positive and negative entries in the second eigenvector $\mathbf{q}_2$ relative to the second eigenvalue $\lambda_2$. $q_2$ and  $\lambda_2$ are also called Fiedler vector and Fiedler value, respectively. 

We use modularity \cite{newman2006modularity} for identifying the polarization of users in the social bulk in two groups, labelled here by $c_1$ and $c_2$: 
\begin{eqnarray}
\label{eq:polarization}
\Phi_{F} = \frac{1}{2m}\sum_{ij} \left[ A_{ij} - \frac{s_i s_j}{2s} \right] \delta_{c_{i}, c_{j}}.
\end{eqnarray}
Here, $A_{ij}$ is 0 is users $i$ and $j$ did not interact, otherwise it is equal to the number of their interactions; $s_i$ indicates the total number of interactions involving the $i$-th user, i.e. its strength, while $s$ is the total number of interactions in the network. Polarization values $\Phi_{F}$ close to 0 indicate no antagonism between opposing factions, while $\Phi_{F}$ close to 1 is relative to strongly opposing factions.

We use the generalization of modularity to more than two groups for establishing the fragmentation of users in antagonizing social groups. The mathematical definition is similar to Eq.~(\ref{eq:polarization}), except for the fact that we consider more possible partitioning into a number communities ($c_{1}, c_{2}, ...,c_{M}$) larger than 2. The number $M$ of existing communities is not known \emph{a priori} and an optimization process must be employed to discover best partitioning of the system. 

We measure social segregation by considering the average size of connected components weighted by the number of their links. Indicating with $\mathcal{C}$ the set of connected components and with $\mathcal{C}_i$ the set of $n_i$ nodes connected by $e_i$ edges in the $i$-th connected component, we define \textit{social segregation} as:

\begin{eqnarray}
\Sigma = 1 - \frac{\sum\limits_{i=1}^{|\mathcal{C}|} n_{i} e_{i} }{\sum\limits_{i=1}^{|\mathcal{C}|} n_i\times \sum\limits_{i=1}^{|\mathcal{C}|} e_i}.
\end{eqnarray}

$\Sigma$ ranges between 0 (a network with a single connected component) and 1 (a network of isolated nodes with no links).

We define infiltration of a given type of users in a given social group $i$ as the fraction of users of that type in group $i$, namely:
\begin{eqnarray}
\mathcal{I}_{s} = \sum\limits_{i=1}^{\mathcal{M}_i} \frac{s_{i}}{u_{i}}
\end{eqnarray}
where $\mathcal{M}$ is the number of groups, $s_{i}$ is the number of accounts of class $s$ in the $i$-th group and $u_{i}$ is the number of users in that group.

\heading{Testing the role of news media.} In the analysis of the social bulk we identified two communities corresponding to news media accounts. In order to test for the influence of these information hubs on human-bot interactions, we performed a test in which we checked the robustness of our results when all users in the above two communities identifying news media accounts were not considered. The removal of news media accounts led to negligible fluctuations (around 0.02) in the fractions of human-bot interactions (cfr. Fig.~1~(a)) and in the total volume of tweets produced by bots (around 0.4\%). These results indicate that a prominent amount of human-bot interactions does not involve news media accounts and it is not influenced by the presence of information hubs.

\bibliographystyle{unsrt}
\bibliography{biblio}

\section*{Acknowledgements}

We acknowledge Pierluigi Sacco for insightful discussion.

\section*{Author contributions statement}

MS, MC and MDD conceived the experiments, MS and MC conducted the experiments, MS and MDD analysed the results. All authors reviewed the manuscript. 

\end{document}